# Sensors in Viticulture: Functions, Benefits, and Data-Driven Insights

Milan Milenkovic,

IoTsense, California, USA

milan@iotsense.com

## Abstract

Use of sensors and related analytical predictions can be a powerful tool in providing data-informed input to a viticulturalist's decision process, complementing their vineyard observations, experience, and intuition. Up-to-date measurements, predictions, and alerts offer actionable insights and suggestions for managing key vineyard operations, such as irrigation, disease control, canopy management, and harvest timing. In many cases, anticipatory interventions can mitigate risks before problems become apparent. Sensor data platforms facilitate implementation of the principles of precision viticulture – doing the right thing, at the right time, in the right place. By offering guidance on the targeting, timing, and dosage of vineyard practices, they can enhance operational effectiveness and efficiency while conserving labor and resources.

This paper is primarily intended for viticulturalists and vineyard managers as a succinct summary of viticulture sensor data platforms, their functions and benefits, how they work, and some practical considerations related to their acquisition and deployment. It may also be of interest to researchers in agriculture and IoT systems.

## Sensor Data Platforms: Functions and Benefits

Use of sensors in viticulture can provide instantaneous and continuous information on vineyard status and predictions to provide data-based guidance for informing a viticulturalist's decision-making process. Commonly used sensors can measure ambient, soil, and plant conditions. Their data are acquired and processed by software that provides visualizations, alerts, and predictions of interest to viticulturalists - such as estimate of crop stages and harvest time, timing of disease management, irrigation, fertilization.

In a typical setting one or more weather stations are installed in the vineyard. Those include an array of ambient sensors such as temperature, relative humidity, precipitation (rainfall), barometric pressure, wind speed and direction. In addition to sensors, the weather station complex includes a processing unit to acquire data. It also contains communication devices to connect to other stations and to the back-end processing complex. Adjunct sensors, such as soil moisture, leaf wetness, cameras and spectrometers can be attached to the weather stations. Alternatively, they can be placed standalone equipped with communications devices to form a local wireless network in the field. At least one of the nodes in the system needs to have the capability for long-haul communication. This enables connectivity to servers that aggregate and archive acquired data to make them available for algorithmic processing and user visualization and interaction.

The software that operates all this consists of the data acquisition part, archival storage, data processing including prediction algorithms often with the aid of artificial intelligence, and the user interface part that provides applications for visualization, interactions, and customizations. The data platform's components reside in three main areas: the in-field gateway, local or remote servers for archival storage and more demanding computations, and user-facing applications on mobile devices and personal computers. More technical details on its operation and components are provide later in this paper. We begin with a description of its functionality and benefits provided to viticulturalists.

Based on the sensor data and algorithmic post processing, the following functions are typically provided by the sensor systems described above:

- Environmental data and weather forecasts
- Tracking of crop growth stages and estimate of harvest time
- Disease alerting and management – prediction of fungal disease and insect growth
- Irrigation recommendations - when to irrigate and how much
- Fertilizing recommendations
- Frost prediction – when to expect and how to mitigate
- Crop yield prediction
- Safety of workers – lighting tracking and prediction, UV light, heat index
- Long term storage and archival data on specific vineyard and blocks

## Environmental data and weather forecasts

The set of sensors dedicated to ambient conditions above ground includes some combination of temperature, humidity, barometric pressure, wind speed and direction, precipitation, solar radiation, and leaf wetness. A GPS sensor may be added for precise location of sensors on the map, which is useful especially when several weather stations are installed in a vineyard.

Depending on the sensor availability, the following readings are typically collected by sensor data platforms

- Temperature
- Relative humidity
- Barometric pressure
- Solar radiation
- Wind speed and direction
- Precipitation (rainfall)
- Leaf wetness

Other above ground common sensors include RGB cameras to remote visual provide visual crop monitoring, and spectral cameras to help assess vine health and vigor. They are described later in this paper.

In addition, short- and medium-term weather forecasts for the vineyard are displayed. They are provided as a part of the service and are usually obtained from commercial services offering specific localized forecasts for the vineyard's specific location. They provide short and medium weather forecast to help manage vineyard activities, such as the timing of spraying or harvesting to avoid windy or rainy conditions. Alternatively, weather data can be obtained from public sources. These are generally less precise because they cover wider areas.

Commonly calculated values that are also reported and used in predictions include

- Dew point
- Evapotranspiration
- Growth Degree Days (GDD)
- Chill hours, chilling units

The *dew point* represents the temperature at which air becomes saturated with water vapor and begins to condense into dew. It is computed based on the temperature and relative humidity.

*Evapotranspiration* is the measure of combined loss of water from the vineyard through evaporation from the soil and transpiration from the grapevines. It is calculated using temperature, solar radiation, humidity and wind speed. There are also specialized evapotranspiration gauges (sensors).

*Growth Degree Days* (GDD) is a cumulative measure used to track heat accumulation and predict the vine growth stages throughout a growing season. It is calculated using average daily temperature minus base temperature that varies slightly for different grape varieties and geographies.

*Chill hours* count the number of hours when temperatures in the vineyard fall between $0^\circ$C and $7^\circ$C (32-45$^\circ$F). Chilling units, such as the Utah Chilling Model, assign different values to hours spent at various temperatures. These factors are used to assess whether vineyard has experienced sufficient cold temperatures for proper vine development in the following season.

Sensor measurements are usually taken in regular intervals, such as 15 min. They are time stamped, possibly annotated with the type and point of origin, and forwarded to be reported and stored. These sensors indicate ambient conditions in the vineyard and are also used to calculate predictions and generate alerts as appropriate. Data can be viewed on user visualization devices, such as mobile phones and personal computers. The readings are displayed in a variety of ways including instant values, minimums, maximums sometimes annotated with times of their occurrence, daily and seasonal averages. Additional visualizations are provided in the form of graphs of historical data that can cover ranges of interest all the way to the beginning of season. Graphs can usually be customized to include components of interest, such as daily and nightly temperatures, humidity, and precipitation. They can often be overlaid with other measurements, such as rainfall and humidity, to facilitate comparisons and correlations of interest.

User visualization systems also include notifications, often configurable, and alerts to conditions that may warrant immediate attention, such as forthcoming inclement weather or buildup of disease pressure.

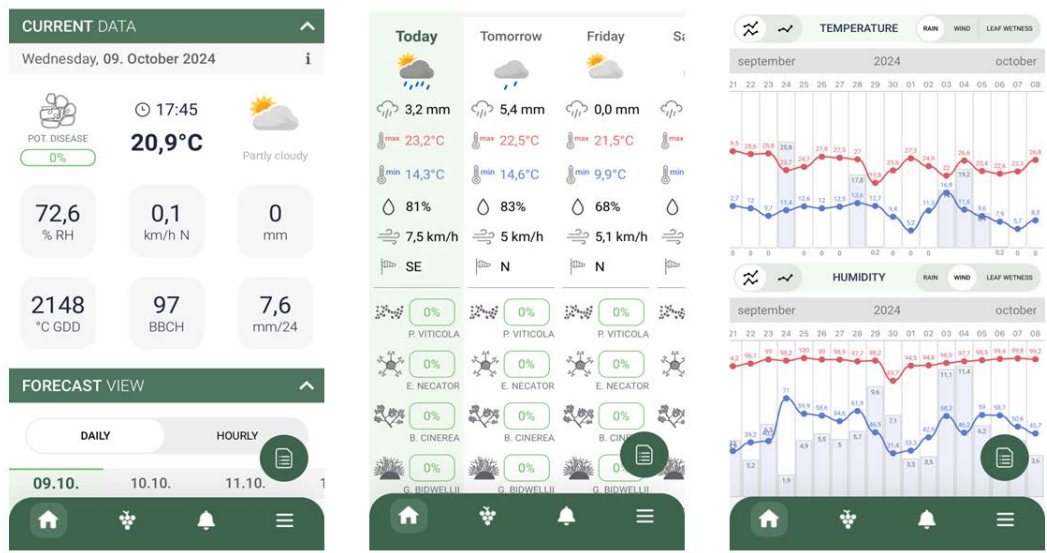

*Figure 1. Viticulture Application Screenshots*

Figure 1 illustrates a representative style of data visualization on mobile devices where the emphasis is on immediacy and readability. It includes three screen shots from a commercial precision viticulture software application. The one on the left is the main/home screen, the other two show weather forecast and pest pressure prediction, and temperature and humidity graphs for a number of days in a sliding window. In this example they are selected to be overlaid with rain and wind measurements, respectively.

In a typical setting, the vineyard where weather stations are installed is outlined on a map provided by a GIS, such as an aerial observation or mapping software. Locations of sensors and stations can be marked manually or automatically in installation where GPS sensors are available. These views can be displayed by application to indicate visually locations of the specific data sources and crop stages and health when supported by sensor measurements such as spectral cameras and NDVI.

More detailed numerical data can be provided on devices with larger screens, such as laptops. One such example is provided by the network of 21 weather stations maintained by San Luis Obispo County in and around Paso Robles, CA wineries. They provide a treasure trove of numerical data for each station, including temperature (daily max, min and time of each), precipitation (daily, monthly, season), wind speed (daily max and time), dew point, relative humidity (max, min), daily solar radiation, evapotranspiration (daily and weekly), GDD (daily, seasonal), chilling, chill hours. The stations also provide data from their elevated temperature sensors mounted on 30 ft (9.144 m) poles to indicate temperature inversion which is useful for frost prediction and mitigation. In addition, many data platforms support creation of customized reports for various data ranges, such as the one illustrated in Figure 2. That depicts temperature, wind, and evapotranspiration data in one of the SLO stations for an earlier year.

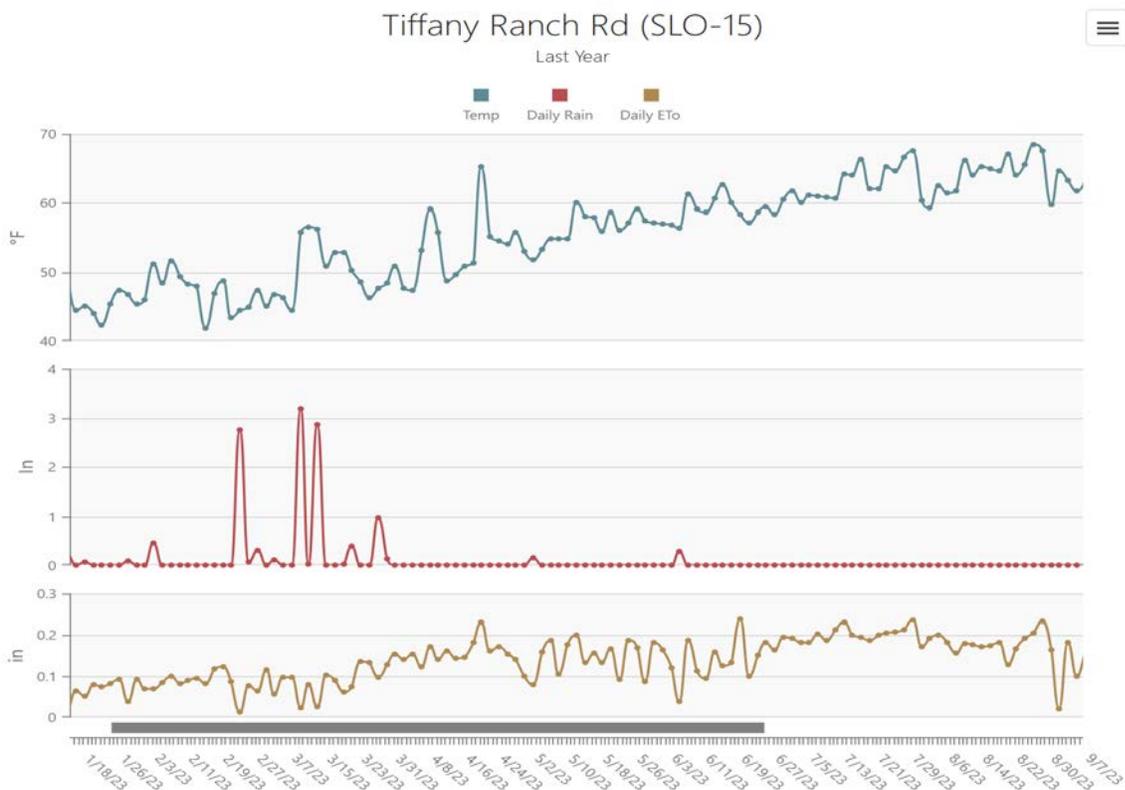

*Figure 2 Customized data report graph*

These figures are an illustration of the variety and range of options in presenting data acquired and processed by viticulture sensor data platforms. They are usually organized in related groupings of information of interest to viticulturalists. Additional data can be viewed on demand in various forms that are customizable by the user. Presence of archival data when the system is in operation for multiple years is useful for comparisons and longer-term trend analyses, inferences, and predictions.

## Tracking of crop growth stages and estimate of harvest time

An important function of viticulture sensor data platforms is tracking and prediction of grape growth stages, from dormancy to harvesting. A primary indicator of the crop growth stages is the Growth Degree Days (GDD). It tracks periods of time when temperatures are suitable for growth. GDD is calculated based on the daily maximal and minimal temperatures, and it is accumulated for the season. Research and observations have provided tables of correspondence of GDD values to the various stages of crop development, including budburst, full flowering, veraison, and harvest. Values can vary based on grape varieties, vineyard management policies, microclimate, and geography. Customized estimates of GDD range correspondence to growth stages have been compiled based on research and observations. Table 1 below provides and illustrative example.

| Crop Stage | Cab Sauvignon, Napa | Usually occurs | Chardonnay, Burgundy | Usually occurs |
|---|---|---|---|---|
| Bud break | 50-100 | March, April | 50-75 | April |
| Flowering | 350-400 | May | 300-350 | May, June |
| Veraison | 1100-1300 | July | 900-1100 | July, Aug |
| Harvest | 2200-2500 | Sep - Oct | 1800-2000 | September |

*Table 1 Example of GDD ranges for crop growth stages*

A related indicator used in industry is the more comprehensive BBCH (Biologische Bundesanstalt, Bundessortenamt und Chemische Industrie). It describes the phenological development of grapes in detail, starting with winter dormancy and ending with leaf fall. BBCH is descriptive as defined, but it can be correlated to calculated GDD values for approximate prediction.

Viticulturalists can use these predictions as an added input into tracking the growth of the crop and an estimate of the onset of important stages when vineyard actions or tests may be advised. They should be used as an approximation and supplemented by the field observations. Recording actual values can be useful for profiling a specific vineyard to refine future predictions.

## Disease management – prediction of fungal diseases and insect growth

Another important function of viticulture data platforms is prediction of disease pressures. Predictions generally indicate the degree of risk and show when and where viticulturalists should pay special attention. Predictions are made by correlation and inferences from the current environmental conditions, historical data, and pest biology. Prediction algorithms can be fairly accurate because insects are exothermic organisms and ambient conditions significantly affect their life stages, including hatching and growth. Similarly, development of fungal diseases such as powdery and downy mildew, strictly depend on temperature, humidity, and solar radiation.

Temperature and data related to humidity, rainfall, and leaf wetness tend to indicate when conditions are favorable for development of insects and fungal diseases. Solar radiation and wind speed and direction data are also a factor used in making predictions. For example, wind speed and direction can be used to infer areas of potential spread of diseases from their points of origin. Advanced platforms can incorporate GIS-based mapping to visualize the spread within and between vineyard blocks.

Prediction algorithms provided by vendors can be based on scientific research, proprietary, or a combination of the two. They can be refined and fine tuned for the conditions in a specific vineyard by comparing predictions with ground truth, i.e. actual field observations. Additional improvements can be obtained by taking into consideration historical data.

In general, predictions of diseases have proven to be very effective for powdery and downy mildew. They are highly to moderately indicative (in decreasing order) for botrytis, black rot, and phomopsis cane and leaf spot. Predictions and forecasts of insect hatching and development range from effective to highly and moderately indicative (in decreasing order) of grape berry moth, grape leafhopper, mealybugs, vine moth, spyder mites, glassy-winged sharpshooter, phylloxera. These results can be improved, and the prediction window extended by installing and analyzing data from insect and spore traps.

In general, such predictions provide data-informed advice on the potential onset and the intensity of disease pressures in parts of the vineyard. This information can be used to decide when and where to spray and when it may not be necessary to spray. Timely spraying can avert and mitigate diseases before they develop. Skipping of spraying, when possible, conserves labor and materials while improving sustainability and eco friendliness.

### Irrigation recommendations

Based on a combination of sensor inputs and derived indicators, viticulture data platforms can suggest when and where to irrigate as well as when irrigation is not needed. This helps to maintain optimal vine health and avoid unwanted degrees of plant water stress. Indication on when irrigation is not needed is useful for conserving water and labor resources and reducing the unnecessary risk of disease development. Irrigation recommendations can include specific details for vineyards and vineyard blocks, subject to the density of sensor placement. They can include amount of irrigation needed and where and when to apply it.

Irrigation recommendations are primarily based on precipitation data and evapotranspiration, which is influenced by temperature, solar radiation, humidity, and wind speed. These factors measure the daily water loss from the vineyard that may need to be replenished. This information is complemented by the data on the amount of irrigation recently applied. Leaf wetness data, when available, can also contribute to decisions by indicating lingering effects of dew or light rain. Localized weather forecast can modulate irrigation decision based on rainfall expectations. For example, irrigation may be skipped if significant rainfall is expected. The effectiveness and accuracy of irrigation predictions can be improved by strategically installing soil moisture sensors and integrating their readings with platform data streams.

Irrigation can be automated by installing additional sensors and actuators. Those include sensors for pressure and flow measurement, along with actuators to control valves and pumps. Depending on their sophistication, some of those systems can precisely target, time, and pace irrigation dosing for specific vineyard areas or blocks. Automated irrigation systems are usually self-enclosed and operate independently from viticulture data platforms. Interfacing the two systems is useful for exchanging information on schedules and completed irrigation tasks. This data can then be factored into subsequent predictions.

### Fertilizing recommendations

Sensor-based software can provide fertilization advice typically by using a combination of soil and environmental data to assess nutrient needs and to optimize fertilization strategies. The goal is to ensure that grapevines receive the right amount and type of nutrients at the appropriate times to promote healthy growth, optimize grape quality, and prevent nutrient leaching or runoff.

Ambient sensor data provide an indication of factors that can affect the timing and dosage of fertilization. For instance, evapotranspiration combined with precipitation indicates effective water loss from the soil and vines, which allows viticulturalists to infer the rate at which nutrients may be taken or lost from the soil. Heavy or frequent rainfall can wash away surface-applied fertilizers or dilute soil nutrient concentrations, especially in sandy soils. Temperature and solar radiation data are crucial for understanding vine growth stages and rates that affect the corresponding need for nutrients.

Wind speed and direction data conditions can be used to plan the application of foliar fertilizers and other treatments. High winds can cause spray drift, reducing the effectiveness of applied nutrients. This information enables viticulturists to time foliar fertilizer applications when conditions are calm to ensure maximum coverage and uptake.

Fertilization advice can be made more precise by incorporating data from additional soil sensors, including nutrient, moisture, pH, soil temperature, and leaf chlorophyl sensors. Some advanced weather stations include spectrometers to measure the incoming and reflected radiation. These readings are commonly used to calculate NDVI (Normalized Difference Vegetation Index) that can assess vine health and vigor. For instance, an unexpected decrease in NDVI during canopy development could indicate plant stress, plant disease, nutrient deficiency, or need for canopy management. Portable NDVI sensors that can also be mounted on tractors or other farm equipment to measure multiple points in a vineyard. Remote sensing tools, such as aerial imagery from satellites, airplanes, and UAVs (Unmanned Aerial Vehicles) equipped with spectral cameras can calculate NDVI and assess vine health and vigor across large areas.

### Frost prediction

In vineyards situated in the areas where frost may occur, viticulture platforms can provide prediction and mitigation advice. Frost prediction can be based on a combination of inputs, such as localized weather forecast, temperature, barometric pressure, relative humidity, solar radiation, dew point, wind, and historical data.

Frost prediction can generate alerts when frosting is expected. The platform may also be able to suggest which of the available mitigation approaches may be beneficial, such as heating or spraying. Temperature inversion measurements, when available, can indicate when the use of wind turbines may or may not be effective. A significant temperature inversion indicates the presence of warmer air above that wind turbines can circulate to reduce the risk of frost.

### Crop yield prediction

Weather sensor data can provide a useful contribution to crop yield estimates. Those are typically based on various models that also incorporate soil conditions, grapevine growth stages, and past historical data to predict the quantity and quality of grape harvests.

Weather data – primarily temperature, precipitation, solar radiation, and wind speed – combined with derived computations such as GDD and evapotranspiration, provide significant information about conditions during the grape growth and the resulting yield. Additional input can be obtained from aerial surveillance combined with NDVI data to provide information on vine vigor, canopy density, and leaf area index that help estimate the overall health of the vineyard and the likely resulting yield.

This information, combined with as much historical data on yield in similar conditions as possible, can be used by Machine Leaning (ML) and artificial intelligence (AI) algorithms to provide more refined predictions of yield. Moreover, these algorithms usually have the capability for adaptive learning from new data. By comparing predictions with outcomes, they improve accuracy of predictions over time.

The accuracy of yield predictions can be affected when there is high variability in vineyard conditions, such as extreme weather events or disease outbreaks. When these factors are present, they should be accounted for to adjust the predictions accordingly. Even when not perfectly accurate, yield estimation models give valuable insights that help viticulturists plan harvests, allocate resources, and make informed management decisions. Early estimates allow viticulturists to anticipate potential shortfalls or surpluses and adjust marketing strategies accordingly.

### Long term and archival data storage

One of the most valuable benefits of viticulture platforms is the ability to store and analyze multi-year and multi-season data for as long as the system has been in operation. Such data can be used for detailed profiling of the microclimate in specific vineyards and blocks and used to track changes over time. One of the more valuable possibilities is to compare growing seasons and identify similarities to and differences from the previous ones in order to project probable outcomes for the current one. Some sensor data platforms facilitate these comparisons by providing overlays of graph data for the time periods and seasons of interest. More sophisticated platforms perform use computational analytics to provide deeper insights.

In general, more data lends itself to richer and more accurate computational analyses. They may produce quantified comparisons, identify and analyze trends. Some platforms apply machine learning and artificial intelligence to enhance results, create additional insights, and make actionable suggestions. The value of the system increases over time as it accumulates more data. These data can be supplemented for additional insights by combining and cross-referencing them with available regional and global databases of weather and viticulture data.

## How it Works

The three major components of viticulture sensor data platforms are (1) sensors, (2) communications, and (3) back-end data processing complex. The system is completed by pieces of software running on end-user devices, such as mobile phones and laptops, that provide user interface for data visualization and human interaction.

### Sensors

Sensors are transducers that convert measurements of some physical phenomenon, such as temperature, into electrical signals. These signals are then converted into numerical values using units of a measurement system, such as US customary units or metric. They have wired interface to the nearest computing element, such as a weather station or a wireless communications module for stand-alone sensors.

Ambient sensors for use in viticulture - such as temperature, humidity, precipitation, barometric pressure, wind speed and direction – are usually clustered and packaged together in a weather station. In addition, the weather station contains compute and communication elements. The computational part, often referred to as a gateway, is used to acquire sensor data, do preliminary processing, and send them via a communication link to the central data aggregation point. Most weather stations support the connection of additional sensors, such as soil moisture sensors, directly or via wireless connections.

Many viticulture data platforms support the installation of auxiliary weather stations to provide more comprehensive coverage of a vineyard and its blocks. In such systems, one of the weather stations is usually a more capable unit that acts as a master gateway. It connects to auxiliary units via a wireless local area network.

Where available, in-field sensor data are complemented by aerial imagery and spectral camera readings that may be obtained from satellites, UAVs, or aircraft flyovers. Localized weather forecasts provide additional inputs that are processed and reported by the data platform.

There are also some other types of sensors that may be installed in the vineyard to provide data of interest to viticulturalists but are usually not interfaced to weather stations. They include spore traps that require samples to be collected and sent to a lab for analysis for early signs of fungal disease development. Some can even perform spectral analysis of captured samples directly in the field unit and report findings via an application. Insect traps, such as pheromone, sticky and light, can also be used to assess their presence and the level of pressure they may pose.

Additional, less common types of sensors, which tend to operate in isolation, are plant probes and monitors. Examples include dendrometers to measure trunk radius variations and electrophysiology sensors that measure plant bio signals.

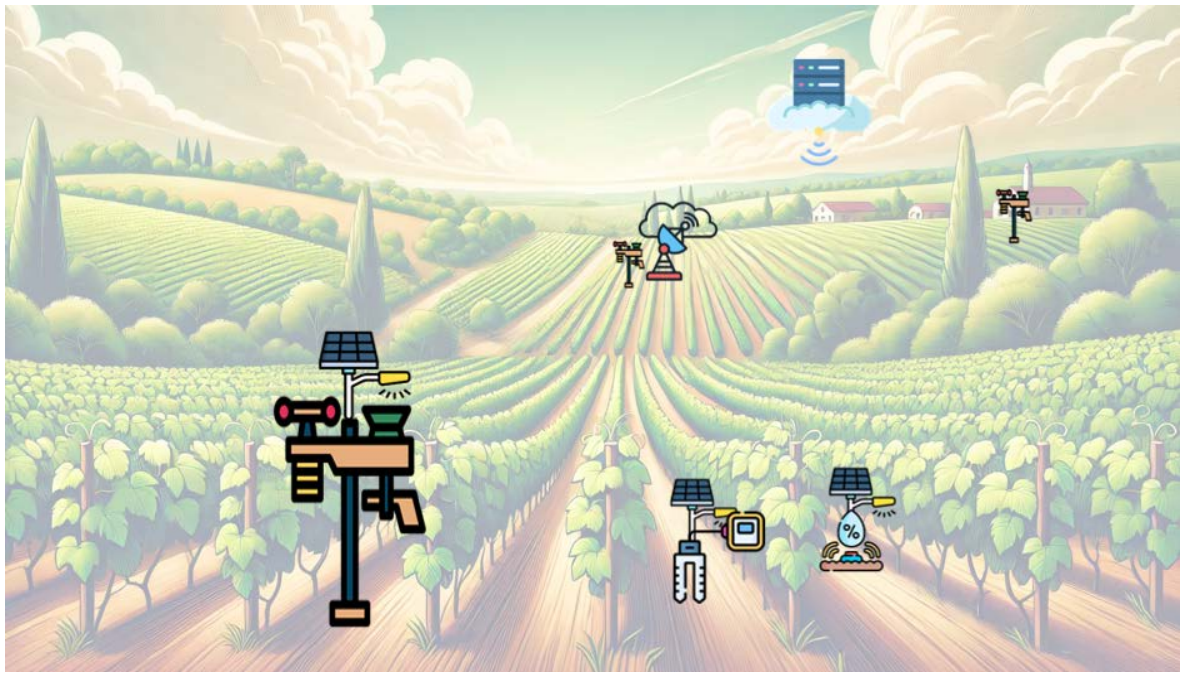

*Figure 3. Vineyard with sensors*

Figure 3 is a stylized representation of a vineyard with three weather stations and soil sensors connected wirelessly. The distant weather station in the center represents a gateway which has wide area network connectivity to the remote platform servers depicted symbolically as if in the cloud. The platform servers would generate reports and drive user interfaces such as those depicted in Figures 1 and 2. Some installations also include RGB cameras for visualizing conditions in the covered part of the vineyard, and spectral cameras for NDVI analysis.

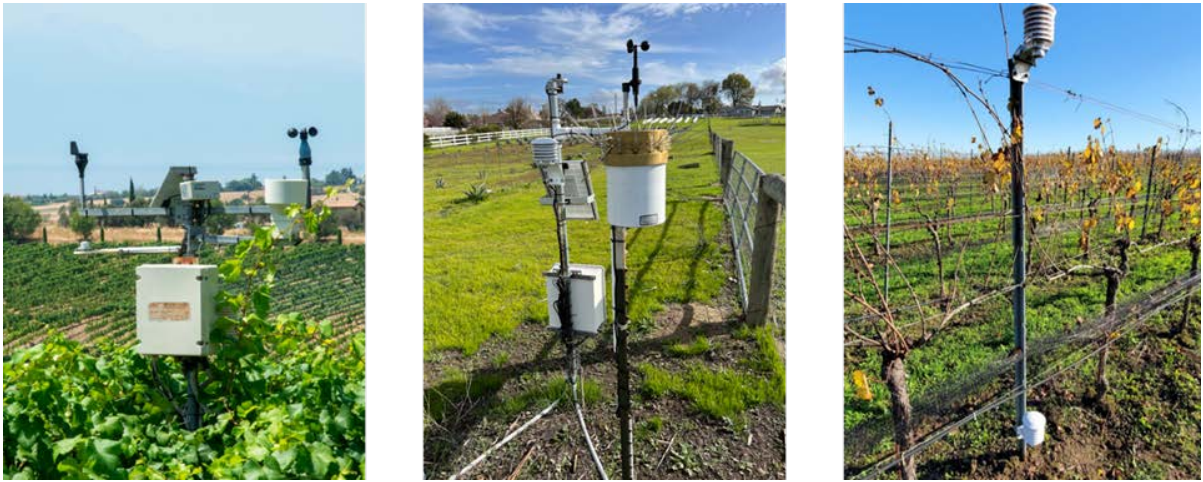

*Figure 4 Weather stations in the field*

## Communications

Gateways in viticulture sensor data platforms aggregate local data collections and communicate them to remote data management components. Those communication links are often long range, such as wide-area networks and cellular connections for vineyards within a coverage distance from a cell tower. Theoretically, with high-gain antennas, the range can extend up to 70 km (43 miles) in rural areas with no obstacles and clear line of sight, significantly less in hilly and populated areas. In addition to legacy GPRS, cellular data standards, such as 4G LTE and 5G, specify versions of lower data rate links optimized for wireless sensors and Internet of Things (IoT) uses – such as LTE-M or NB-IoT. In the areas where those offerings are not available, regular cellular links may also be used to connect gateways, though possibly at higher cost. All variants of cellular offerings operate in the licensed spectrum slices assigned to telcos and their use requires paid subscription.

An alternative is to use unlicensed spectrum which is free but may be more prone to interference from nearby users. In configurations where the viticulture gateway is located within 1 km (3000 ft) of the internet access point, low energy BT (Bluetooth) long range introduced in version 5 may be used. WiFi may also be used if there is coverage within the gateway's range and sufficient power to support it. Other approaches include long-range wireless networks, such as LoRaWAN (Long Range Wide Area Network). LoRa WANs use public radio frequencies in the unlicensed spectrum. However, they require supporting infrastructure, such as edge gateways and connection servers, which must be installed unless already available from a public or commercial source in the range of the vineyard area.

In any case, the gateway must be configured with the necessary components for the chosen network: communication interfaces, radio transceivers, and modems. For commercially provided communications, the gateway must also include an access and subscription module, such as a SIM card for cellular networks. Gateways are usually powered by a combination of solar cells and a rechargeable battery backup.

Systems that support auxiliary weather stations and stand-alone wireless sensor connections also need to form and manage an intra-vineyard wireless network for their communication. These are typically short- to medium-range wireless networks designed to operate in low-power mode for battery operated devices. Radios and modems for such networks, which commonly conform to the IEEE 802.15.4 link layer standard, use unlicensed portions of the spectrum that are free to use. They run communication protocols, which are often proprietary, designed to support the required data rates for power-constrained devices. With proprietary protocols, all communication interfaces for auxiliary devices must be obtained from a single vendor.

Intra-vineyard wireless networks use different topologies, with star and mesh configurations being the most common. In star topologies, auxiliary nodes can communicate directly only with the central node which is the gateway. Mesh configurations support some point-to-point connections between auxiliary nodes. As a result, they can extend the network's overall range, but this comes at the cost of increased complexity and the need to use routing protocols.

## Back-end data processing complex

Data from the vineyard sensors are collated, time stamped and sent to the back-end processing complex. Primary functions of the back end are data storage, retrieval, and execution of prediction algorithms including ML and AI as supported. Some data are processed in flight to generate updates and alerts when the conditions that trigger them are met. Usually all acquired sensor data are stored for archival and post-processing purposes. Stored data are retrieved to generate configured and on-demand user reports, and as inputs to prediction algorithms. Data stored for long-term archival purposes may be sub sampled to conserve storage. Archival data are also used for compliance reports, cross-season comparisons, and to increase the accuracy of prediction algorithms based on ML and AI, which generally work better when large amounts of data are available.

Depending on the installation and data volume, implementations of back-end can range from a single on-premises server to one or more virtual machines operating in the cloud. In general, on-premises servers can provide a high degree of data privacy and, in some cases, a perceived cost advantage. However, management of servers requires IT skills and diligence in keeping backups and maintaining system uptime, which may not be what wineries want to invest in. A cloud-based back end provides several benefits, including a professionally managed service with uptime guarantees, reliable backups, timely software updates, and ubiquitous accessibility for users and UI endpoints. These platforms can be more cost effective, as they can balance the load across multiple users.

## User Interface Modules

Viticulture sensor data platforms provide user interfaces for visualization and data entry as illustrated in Figures 1 and 2. They are used to visualize sensor data reports, historical data of interest, predictions such as pest pressure and growth stages generated by the algorithms. The platforms also provide alerts to indicate need for attention or intervention and accept user inputs such as recording of observations in the field. Some aspects of visual presentations and alerts can be configured to suit user's needs and preferences. Most platforms also support customized data reports for specific variables and data ranges of interest from hours and days to entire seasons and even multiple years.

These interfaces are usually available on mobile devices and personal computers, such as laptops or desktops. They may be implemented as customized applications for specific devices or in internet formats, such as HTML, for presentation via browsers. Browser versions work across multiple devices and can be more convenient to access, but they tend to be a bit more difficult to customize and may not display equally well on all devices. When applications are used, they are downloaded and installed on user devices. Mobile applications are commonly available for download from popular sources, such as Apple App Store and Google Play Store.

Many platforms support multiple users with separate authentication logins. Some provide support for concurrent access by multiple authorized users, which may be desirable in larger wineries. In such cases, it is necessary to maintain data consistency and integrity by coordinating potentially conflicting updates. Supporting platforms typically achieve this by creating different access privileges to specific user classes, some of which may be read only.

## Practical Considerations

There are many practical considerations involved in selecting and deploying a viticulture data platform in a vineyard. Some of the more important ones include scope and coverage, capabilities and features, placement and density of weather stations and sensors, data formats and interoperability, buy vs lease, and maintenance.

A data platform needs to be comprehensive in its sensor coverage and be able to work with supplemental third-party sensors of interest for a particular vineyard. Basic sensors include temperature, humidity, wind speed and direction, precipitation, solar radiation and support for leaf wetness sensors. RGB cameras may be provided for remote visual inspection, and spectral and infra-red cameras for plant and crop health assessment. It is a common practice to also provide hardware and software interfaces for optional third-party sensors such as soil moisture and temperature. In installations where multiple weather stations are to be used, it is preferable to choose hardware solution that supports inter-vineyard wireless networking. Otherwise, each station may need to connect directly to the long-haul network, which increases the cost and complexity.

The placement and density of weather stations and sensors depend on the vineyard management objectives, size, and topography. In general, the recommended density for weather stations in vineyards is between one weather station per 10 to 20 acres (4 to 8 hectares) in areas with uniform conditions to about one per 5 acres (2 hectares) or less in regions with significant microclimate variations. Weather stations should be placed strategically in representative zones based on elevation, exposure, and vineyard block structure. Similar considerations apply to leaf wetness and soil sensors when used. Weather stations and sensors are usually installed by the hardware vendor or its authorized representatives. Some vendors provide "plug and go" prepackaged solutions, often tripod based, that users can install in simpler configurations.

Some hardware vendors of viticulture weather stations bundle their offering with software, for which they usually charge a recurring fee. In some cases, software for the data platform may be obtained from third parties. This is possible if the hardware vendor publishes their sensor data APIs for the software to use. The tradeoff is that software-only providers often deliver more advanced functionality and a richer user interface, but this must be balanced against the potentially higher overall system costs. Additional challenges can arise from dealing with multiple vendors if operational issues occur.

Viticulture data platforms generally provide data visualizations, alerts, and various degrees of prediction. Many are increasingly claiming the use of ML and AI for this purpose, but what matters to viticulturalists are the types of predictions available and their accuracy. This is difficult to ascertain without the actual operational experience and comparison of predictions with outcomes. In practice, sensor data reports and predictions should be treated as a data-based advisory input to indicate what may need attention or intervention in the vineyard.

One of the key benefits of using viticulture data platforms is the accumulation and processing of long-term data. Proper data management includes resiliency and ability to retain access in the presence of hardware and software failures. This is accomplished through redundancy, replication, and backups, all of which are easier to implement in cloud-based implementations. It is important for viticulturalists and vineyard managers to own their data. In addition to privacy this means the ability to access data even after vendor changes. The best way to accomplish this is to have the platform vendor document formats of the data kept in storage in a manner that can be retrieved by some other applications should that become necessary. While highly desirable, these objectives can be difficult to accomplish for small wineries without IT expertise and staff. However, they should enquire what kind of guarantee can the software vendor provide for maintaining the safety, integrity, longevity, and accessibility of vineyard data.

One other consideration is whether the viticulture sensor data platform needs to be able to interface with other systems that may be used or are of interest to vineyard managers, such as irrigation automation, asset tracking, and workflow management. For example, this can result in being able to create irrigation activation schedules based on predictions, and to use the irrigation completion data as inputs for more comprehensive reports and predictions. In general, such connections can increase efficiency and streamline vineyard operations when conforming systems are being used.

Acquisition of viticulture data platforms can take several forms. Besides the traditional buy and own, they range from purchasing hardware with licensing of software, often on a recurring basis, to platform as a service where the vendor installs the hardware and operates the software for a recurring monthly or annual fee. The latter provides for both hardware and software maintenance which otherwise can induce additional cost and complexity in the traditional arrangements. Some vendors are also experimenting with outcome-based pricing, but those are not yet common.

## Summary and conclusion

Viticulture sensor data platforms integrate field measurements, communications, and computational analytics to provide viticulturalists with actionable information. Their value extends beyond immediate alerts and recommendations: long-term operation produces multi-season records that help profile vineyards, enable cross-season comparisons, and quantify how climatic shifts affect local growing conditions.

Selecting and deploying a platform requires weighing factors such as sensor coverage, placement density, interoperability, and costs of ownership versus subscription. Given the critical role of data, it is also important to plan for resilience, portability, and continued accessibility even if vendors or technologies change.

Used effectively, these platforms can enhance vineyard decision making by adding a data-informed perspective to complement field observations and intuition. Their benefits lie not only in immediate operational guidance—such as irrigation, canopy management, disease and pest control—but also in building a cumulative dataset that improves predictions and supports long-term vineyard strategy.

## Acknowledgments

The author acknowledges valuable exchanges with vendors and practitioners during several major industry events, including the WIN Expo (Santa Rosa, CA) and the Unified Wine Symposium (Sacramento, CA) in 2023-2025. In addition, private discussions with viticulturalists and industry experts provided further insights that informed this paper. Special thanks to Alexandra Papadaki (Αλεξάνδρα Παπαδάκη) and Stergios Giannos (Στέργιος Γιάννος) of Ktima Gerovassiliou (Epanomi, Greece), Vukasin Pejovic (Atfield), and Mark Battany (independent consultant, San Luis Obispo, CA). Affiliations are provided for identification only, and the views expressed in this paper are solely those of the author.

## Terms of Use